\documentclass{iaus}
\usepackage{graphicx}
\usepackage{epsfig}

\title[Magnetic field of 16 Peg B3 V] 
{The magnetic field of the B3V star 16 Pegasi}

\author[Henrichs, Neiner, Schnerr, Verdugo et al.]   
{H.F. Henrichs$^1$%
,
C. Neiner$^2$, R.S. Schnerr$^3$, E. Verdugo$^4$, A. Alecian$^5$,
C. Catala$^6$, F. Cochard$^7$, J. Guti\'{e}rrez$^2$, A.-L. Huat$^2$,\\ J.~Silvester$^5$, \and O. Thizy$^7$}

\affiliation{$^1$Astronomical Institute, University of Amsterdam, Amsterdam, Netherlands,
$^2$GEPI, Observatoire de Paris, CNRS, Universit\'e Paris Diderot; 5 place Jules Janssen, 92190 Meudon, France, 
$^3$Inst.\ for Solar Physics, Royal Swedish Academy of Sciences,
Stockholm, Sweden, 
$^4$European Space Astronomy Centre (ESAC), Madrid, Spain, 
$^5$Dept. of Physics, Royal Military College of Canada, Kingston, Canada, 
$^6$LESIA, Observatoire de Paris, CNRS, Universit\'{e} Paris Diderot; 5 place Jules Janssen, 92190 Meudon, France, 

$^7$Shelyak Instruments, Revel, France
}
\pubyear{2009}
\volume{259}  
\pagerange{100--100}
\date{November 2008 and in revised form ??}
\jname{Cosmic Magnetic Fields: From Planets, to Stars and Galaxies}
\editors{K.G. Strassmeier, A.G. Kosovichev \& J.E. Beckman, eds.}
\graphicspath{{converted_graphics/}}
\begin{document}

\maketitle

\begin{abstract}
The Slowly Pulsating B3V star 16 Pegasi was discovered by \cite[Hubrig \etal\ (2006)]{Hubrig06} to be magnetic, based on low-resolution spectropolarimetric observations with FORS1 at the VLT. We have confirmed the presence of a magnetic field with new measurements with the spectropolarimeters Narval at TBL, France and Espadons at CFHT, Hawaii during 2007.
The most likely period is about 1.44 d for the modulation of the field, but this could not   
be firmly established with the available data set. No variability has been found in the UV stellar wind lines. Although the star was reported once to show H$\alpha$ in emission, there exists at present no confirmation that the star is a Be star.

\keywords{stars: magnetic fields, techniques: polarimetric, stars: atmospheres, stars: individual (16 Peg), stars: early-type, stars: winds, outflows, stars: rotation, stars: pulsations}
\end{abstract}

\section{Introduction}

The B3V star 16 Peg received our particular attention because more than 50 years ago it was reported at a single occasion to show emission in H$\alpha$, giving the star its Be status. However, this emission was never observed again, also not in our data set, and hence this star is probably not a Be star.\\
We describe the magnetic analysis leading to the confirmation of the presence of a magnetic field. 
We also summarize the period search in these data, and conclude that the most likely rotation period is near 1.44 d, but the data set is actually too poor to firmly constrain the period.

\section{Magnetic Analysis}
We applied the least-squares deconvolution method (LSD, \cite[Donati \etal\ 1997)]{donati97} to the spectral lines sensitive to magnetic effects in the 38 obtained Stokes V spectra (see {\sl e.g.} Fig.~1 left), yielding the longitudinal component of the field, averaged over the star, $B_{\ell}$.  In Fig.~1 (right) the magnetic data are plotted as a function of time, including the two FORS1 data points from \cite[Hubrig \etal\ (2006)]{Hubrig06}. We found a varying magnetic field, with absolute maximum values of about 170 $\pm$ 45 G. The smallest error bars are about 30 G. The results are rather robust against integration limits in the Zeeman signature, but we note that these are preliminary in the sense that the used linelist can still be optimized.

\begin{figure}[htp]
\centering
\leftline{\includegraphics[bb=30 333 514 801,height=4.6cm,keepaspectratio]{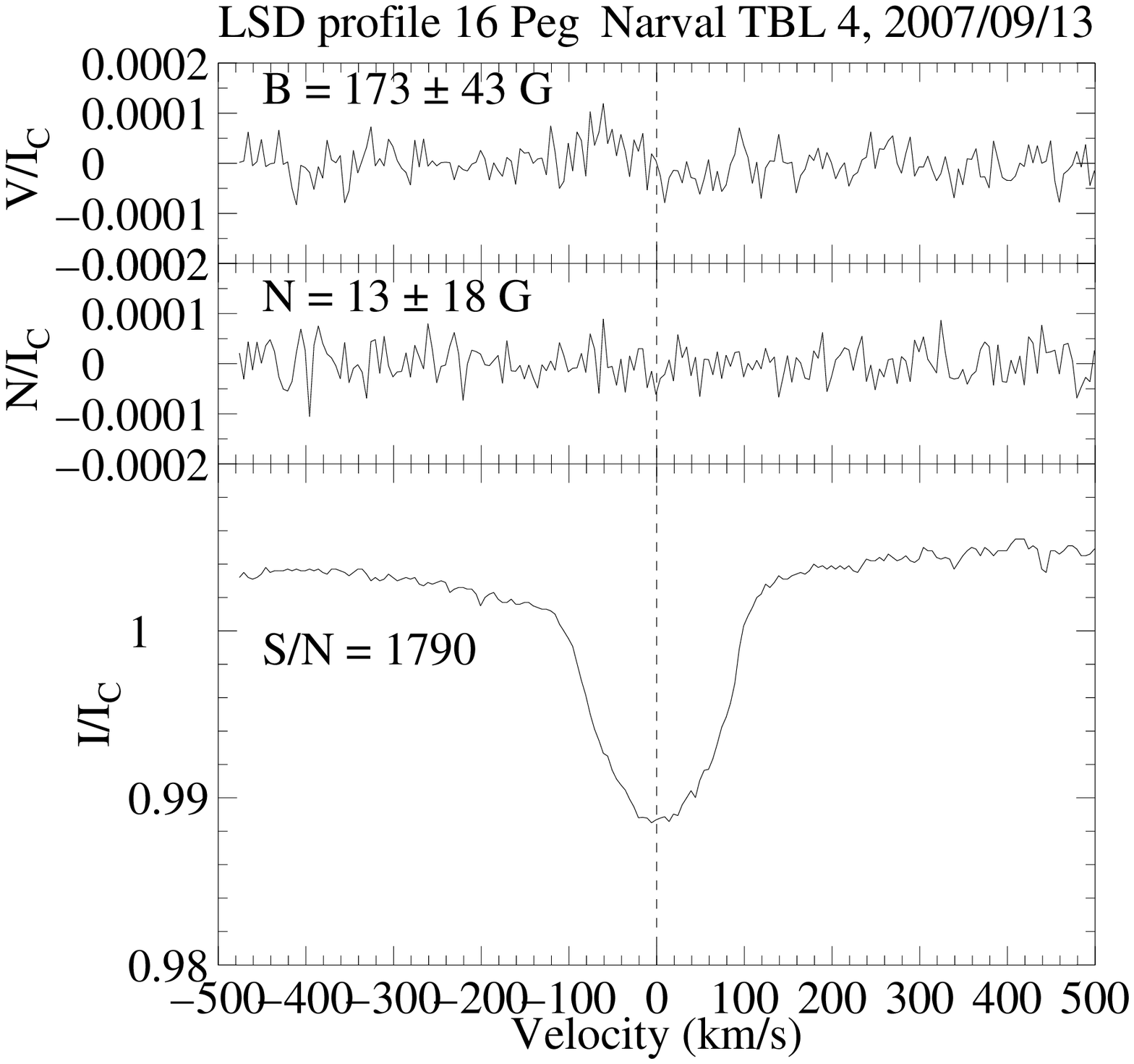}}
\hfill
\vspace{-4.9cm}
\rightline{\includegraphics[bb=0 0 300 214,height=4.6cm,keepaspectratio]{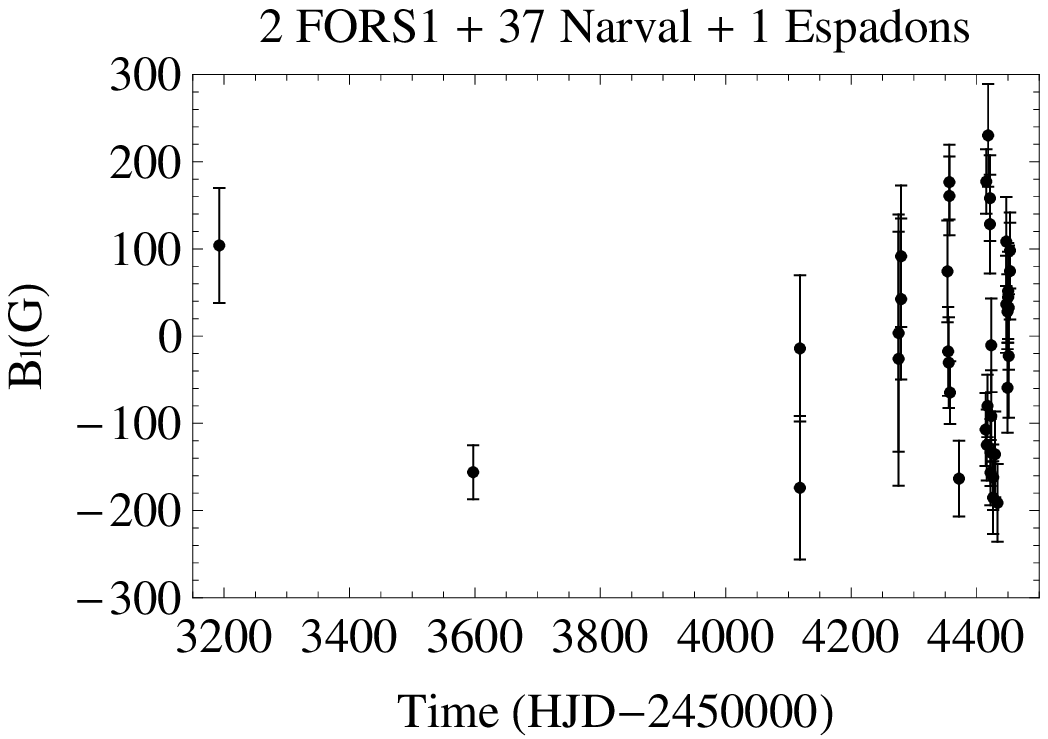}}
\caption{{\sl Left:} Example of a Stokes $V$ Zeeman signature. Top to bottom: LSD  circularly polarised Stokes $V$, (null) $N$, and  unpolarised $I$ profiles of a Narval spectrum. 
 {\sl Right:} Magnetic data history of 16 Peg, including the first 2 discovery datapoints with FORS1 by Hubrig et al.
}
\end{figure}

\begin{figure}[htp]
\centering
\leftline{\includegraphics[bb=63 454 528 753,height=4.3cm,keepaspectratio]{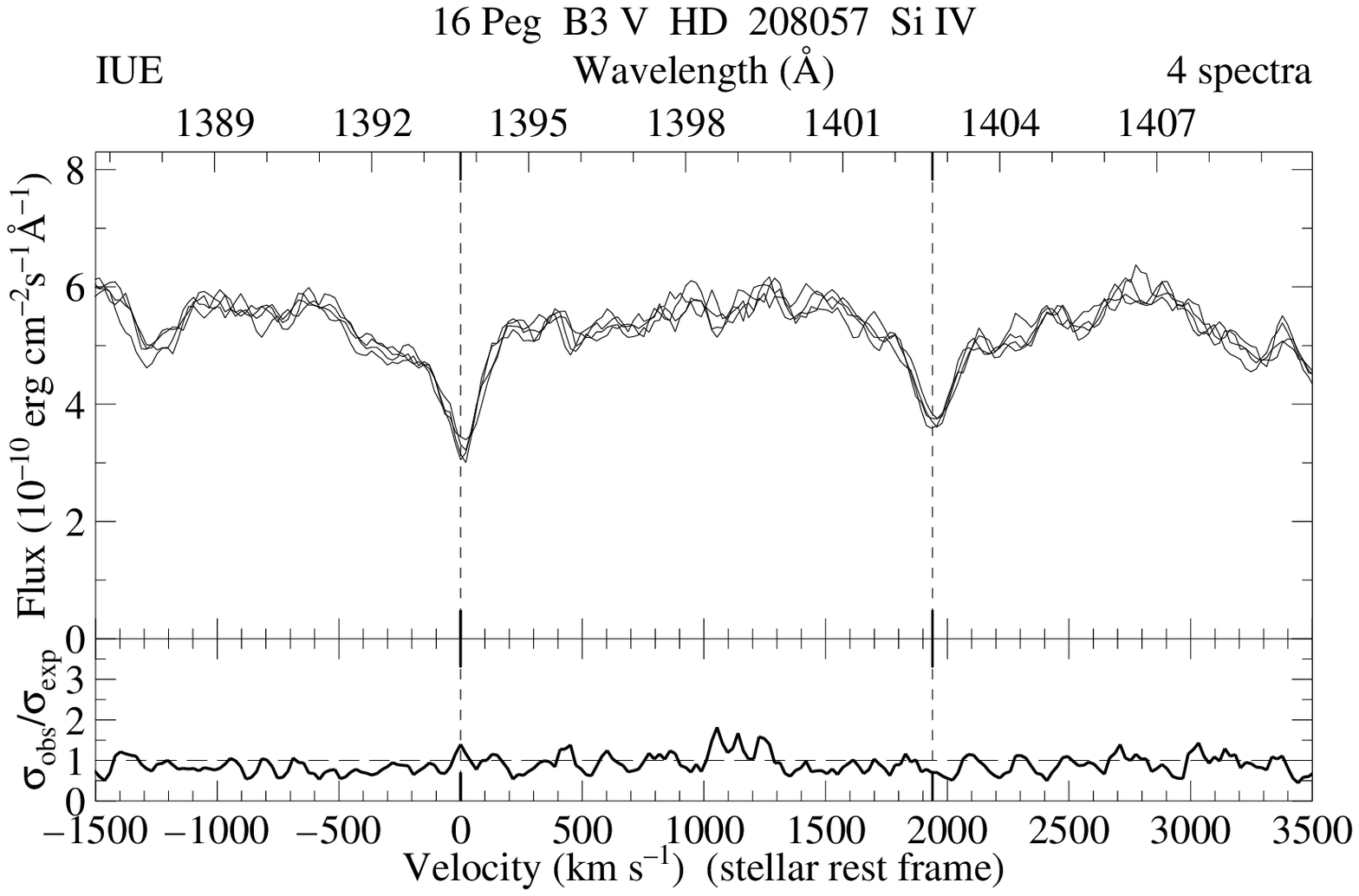}}
\hfill
\vspace{-4.7cm}
\rightline{\includegraphics[bb=0 0 300 213,height=4.3cm,keepaspectratio]{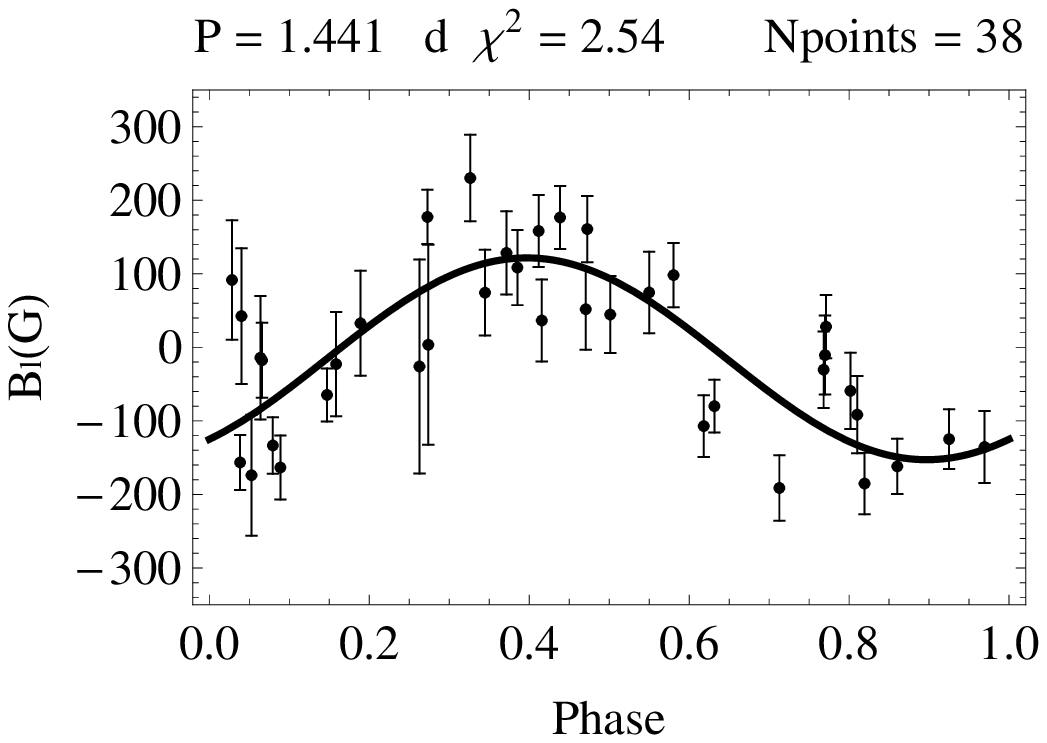}}
\caption{{\sl Left:} Overplot of ultraviolet Si IV resonance lines of 16 Peg. The significance of the variability, expressed as the square root of the ratio of the measured to the expected variances in the lower panel shows no significant variability.
 {\sl Right:} Phase plot of Narval + Espadons magnetic data overplotted with the best weighted fit period of 1.441 day.
}
\end{figure}

\section{Period search and conclusions}
We analysed the 5 available high-resolution UV spectra, taken over 9 years and searched for variability in stellar wind lines. Although such a variability is expected for magnetic B stars, no variability at the S/N level of about 15 could be found (see Fig.~2 left). 

A weighted sinusoidal fit to the data with starting values taken at maximum power of a (unweighted) CLEAN analysis resulted for the 2007 data in $P$ = 1.441 d with a reduced $\chi^2$ = 2.5. Including the two earlier FORS1 points changed the fit significantly, showing the poor coverage and varying quality of the data set.  This period is consistent with $P_{\rm rot}$/sin$i$ = 1.6$\pm$0.6 d from the model stellar parameters given by \cite[Hubrig \etal \ (2006)]{Hubrig06}: $v$sin$i$ = 104$\pm$6 km/s and  R/$R_{\odot}$ 3.2$\pm$1.0, which constrains the inclination angle.

\cite[De Cat \etal\ (2007)]{DeCat07} found three photometric periods in this SPB star: 0.8905 d,  1.24668 d and  0.4039 d. Our proposed magnetic period is not consistent with a rotational splitting.

New spectropolarimetric data with a proper coverage of the suggested rotation period is obviously needed.

\end{document}